\documentclass[aps,pre,twocolumn,groupedaddress,showpacs]{revtex4}

\usepackage{graphicx,amssymb}

\newcommand{\dd}{{\rm d}}

\begin{document}

\title{Structural transitions of monoolein bicontinuous cubic phase induced by inclusion of protein lysozyme solutions}

\author{S. Tanaka}
\affiliation{Faculty of Integrated Arts and Sciences, Hiroshima University, 1-7-1 Kagamiyama, Higashi-Hiroshima 739-8521, Japan}    
\email{shinpei@hiroshima-u.ac.jp}

\author{S. Maki}
\author{M. Ataka}
\affiliation{National Institute of Advanced Industrial Science and Technology (AIST, Kansai), 1-8-31 Midorigaoka, Ikeda 563-8577, Japan}

\begin{abstract}   
Inclusion of protein lysozyme molecules in lipidic monoolein cubic phase induces a transition from a $\rm Pn\bar{3}m$ structure to $\rm Im\bar{3}m$ one. Small-angle X-ray scattering (SAXS) method with high intensity synchrotron radiation enabled us to follow closely the transition depending on the conditions of lysozyme solutions. We showed that concentrated lysozyme solutions induced the appearance of the $\rm Im\bar{3}m$ structure coexisting with the $\rm Pn\bar{3}m$ structure. From the relation between the lattice parameters of these two structures it was shown that they were related by the Bonnet transformation of underlying triply periodic minimal surfaces. We found that the transition also occurred at lower lysozyme concentration when NaCl induced attraction between lysozyme molecules. The origin of the transition was considered as a frustration in the cubic phase where lysozyme molecules were highly confined. A simple estimation of the frustration was given, which took into account of the translational entropy of lysozyme molecules. At the highest concentration of lysozyme and NaCl the $\rm Im\bar{3}m$ structure was found to disappear and left only the $\rm Pn\bar{3}m$ structure. This was probably either due to the crystallization or phase separation of lysozyme solutions ongoing microscopically, which absorbed lysozyme molecules from channels of the cubic phase and thus removed the frustration.

\end{abstract} 


\maketitle

\section{Introduction}
The lipidic cubic phases attract many scientists' interests because of their complex and yet highly symmetric structure. Microscopically, bilayer membranes of lipid molecules are in a disordered liquid state, whereas mesoscopically the membranes form various crystallographic structures. The relevance of the lipidic cubic phase to diverse fields of sciences and technologies includes the membrane fusion of living cell \cite{Caffrey03} and drug delivery system using nano-particles of the cubic phase \cite{Larsson05}. 

Recently, as one of the applications of the lipidic cubic phase, crystallization of {\em membrane} proteins \cite{Landau96} was conducted in a cubic phase of lipid monoolein. The idea is that the lipidic cubic phase provides membrane proteins with an environment as natural as the one on/in biological membranes. The method enabled one to crystallize a membrane protein, which was difficult to crystallize by a conventional method \cite{Landau96}. 

On the other hand, the crystallization of a {\em globular} protein, hen egg-white lysozyme, was found to be possible \cite{Landau97} in the cubic phase and was later found to even be enhanced in the cubic phase \cite{Tanaka04}. Lysozyme molecules reside in a narrow aqueous space between membranes of the cubic phase \cite{Larsson83a,Larsson96}, which is often called channels. This indicates that the lipidic cubic phase works as a nanometer-sized complex yet well-characterized container to accommodate lysozyme solution. The mechanism of the enhanced crystallization has been proposed \cite{Tanaka04}, that is, the frustration caused by the confinement of lysozyme molecules into the channels of the cubic phase generates the additional chemical potential.

The frustration, in turn, should affect the structure of the cubic phase. Razumas et al. \cite{Larsson96} showed that lysozyme inclusion in the cubic phase of monoolein (space group $\rm Pn\bar{3}m$) induced a new cubic phase (space group $\rm Im\bar{3}m$). The mechanism of the structural transition induced by lysozyme, however, is not fully understood yet. Kraineva et al. studied the transition of the monoolein cubic phase induced by the inclusion of cytochrome {\it c} and reported an appearance of a new phase whose symmetry was $\rm P4_332$ \cite{Winter05}. There are no reports so far, at least to our knowledge, that study the structural changes in the cubic phase depending on the crystallization conditions of a protein solution in the channels.

Understanding the structure of the cubic phases is important for making further use of them for protein crystallization, since the environment where protein molecules reside determines the degree of confinement, which is the key to understand the crystallization behaviors in the cubic phase. In this study, therefore, we focus on the structural changes in the cubic phase induced by either crystallizing (supersaturated) lysozyme solutions or not crystallizing (undersaturated) ones. NaCl, the most common salt to induce crystallization of this protein, is used as a crystallization agent to introduce attraction between lysozyme molecules. The cubic phase structures depending on the concentrations of lysozyme and NaCl are determined by small-angle X-ray scattering (SAXS). Utilizing the synchrotron radiation, we could detect fine structures ranging roughly 1-100 nm in size present in a sample volume, which enabled us to follow closely the structural changes depending on the sample conditions.

\section{Experiments}
Monoolein (1-oleoyl-rac-glycerol) was used as purchased from Sigma-Aldrich. Hen egg-white lysozyme (Seikagaku, six times crystallized) was dissolved in 0.05 M(mol/l) sodium acetate buffer. The pH was adjusted at 4.50$\pm$0.05 using a 6 M HCl solution. Lysozyme concentration was determined, after filtration through a 0.1 $\mu$m filter, by UV spectrophotometry (specific absorbance $A_{280}=2.64$ ml/mg cm). We use 10$\times$weight(g)/volume(l) as \% for lysozyme concentration. NaCl solutions of known concentration, whose pH were adjusted at 4.50 by 0.05 M sodium acetate, were then added to lysozyme solutions. The concentrations of lysozyme and NaCl in this study are those of parent solutions into which monoolein was added. 

Monoolein was melted and added into enough amount of solution to ensure that it was fully hydrated. A transparent gel was instantly formed at the mixing, and excess solution around the gel was then wiped off with filter paper. Temperature was kept at 20$^\circ$C by water circulation.   

The solubility of lysozyme crystals in the monoolein cubic phase at $20^\circ$C was studied by the seeding method as follows. Microcrystals of lysozyme ($\sim 10\ \rm \mu m$) made by crushing macro crystals were added to a solution just before it was mixed with monoolein. Observations of the growth or dissolution of the microcrystals with an optical microscope was used to determine the solubility curve, on which neither crystal growth nor dissolution should take place. 

Crystals and other precipitates in the cubic phase were photographed with a bright field microscope as well as with a fluorescence microscope. For fluorescence microscopy, lysozyme molecules were dyed with eosin Y (Sigma-Aldrich). Eosin molecules are reported to form 1:1 complex with lysozyme molecules without inhibition of its enzymatic activity \cite{Jordanides99}. 50 $\mu$l eosin Y solution (0.05\%=7.7$\times 10^{-4}$M) was added to 1 ml lysozyme solution (10\%=$7.0\times 10^{-3}$M) in order to ensure that almost all eosin molecules were bound to lysozyme. The lysozyme solution was then filtered and diluted appropriately. 

SAXS experiments were conducted using a small-angle beamline (BL40B2) of SPring-8 (Japan). The wavelength was 1.0\AA, and the $q$ range used was $q=0.5-2\ {\rm nm^{-1}}$ which covers positions of about 7 diffraction peaks from monoolein cubic phases. Each experiment was done about 12 hours after sample preparation. The exposure time was 20 seconds. 

The lattice parameter $a$ of the cubic phase was determined from a ratio of spacing of peaks using the equation,
\begin{equation}
q_p(h,k,l)=\frac{2\pi}{a}(h^2+k^2+l^2)^{1/2}
\label{lattice}
\end{equation}
where $q_p$ is the peak position and $h$, $k$, and $l$ are the Miller indices of the peak.

\section{Results}
\subsection{Solubility of lysozyme in the cubic phase}
The solubility curve of lysozyme crystals in the monoolein cubic phase is shown with open diamonds with error bars in Fig. \ref{diagram}, which includes the overall results of our SAXS measurements. The location of the solubility was identical within the experimental error to the one we determined kinetically at 18$^\circ$C \cite{Tanaka04}. Crystals were not observed in the samples during SAXS experiments. This means that the samples whose condition was above the solubility were kept in a supersaturated (hence quasi-stable) state during the SAXS experiments.

\subsection{Without lysozyme}

\begin{figure}
\includegraphics[width=5.5cm,angle=-90]{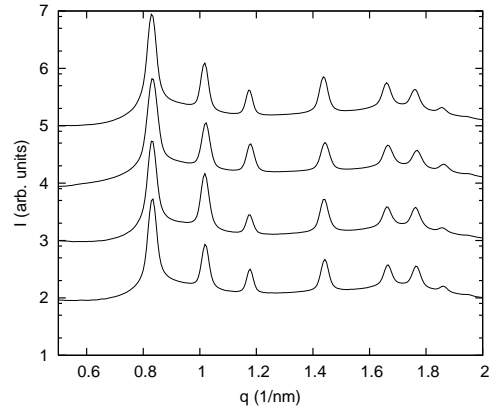}
\caption{SAXS patterns from the cubic phase of monoolein with NaCl in acetic buffers. Lysozyme was not added. NaCl concentration was increased from 0 M(bottom) to 0.3 M(top) by 0.1 M. The baselines of the profiles in this and other figures are shifted to facilitate visualization.  }
\label{lyso0}

\end{figure}

Before studying the effect of introducing lysozyme on the cubic phase, we confirmed that our buffer solutions containing no lysozyme but NaCl and sodium acetate did not affect the structure of the cubic phase. Figure \ref{lyso0} shows SAXS patterns coming from the crystallographic structure of the cubic phase. The solutions in the cubic phase contained NaCl from 0 to 0.3 M and sodium acetate 0.05 M. It is already known that high concentration (more than 1.0 M) of NaCl affects the cubic phase structure \cite{Caffrey01}. As shown in Fig. \ref{lyso0}, however, our buffer solutions containing low concentration of NaCl did not induce any detectable change in the peak positions. The peaks shown in Fig. \ref{lyso0} were relatively narrow, whose full width at half maximum (FWHM) gave us an estimation of the cubic liquid-crystal domain size $L$ using Scherrer formula, 
\begin{equation}
L = 0.9\lambda / B\cos{\theta}
\end{equation}
where $\lambda$ is the wavelength of X-ray, $B$ the FWHM of a peak, and $2\theta$ the scattering angle. We obtained $L\sim 300$ nm irrespective of the sample conditions. 

All the data shown in Fig.~\ref{lyso0} have peaks with a ratio of spacing $\sqrt{2}$, $\sqrt{3}$, $\sqrt{4}$, $\sqrt{6}$, $\sqrt{8}$, $\sqrt{9}$, which are indexed as (110), (111), (200), (211), (220), (221) reflections. The lack of specific reflections indicates that the structure has a symmetry assigned by the space group $\rm Pn\bar{3}m$. The lattice parameters $a$ estimated using eq.~\ref{lattice} are $10.7\pm 0.1$ nm independent of the NaCl concentration. The value of $a$ agrees well with literature \cite{Caffrey00}. These results ensure that the cubic phase structures do not change by the existence of NaCl in the range used in this study (0-0.3 M) as well as by that of sodium acetate (0.05 M). Therefore we can safely assume that any changes that appear later in SAXS patterns must reflect the presence of the  proteins and/or their interactions.

\subsection{Effect of Lysozyme without NaCl}

\begin{figure}
(a)\includegraphics[width=5.5cm,angle=-90]{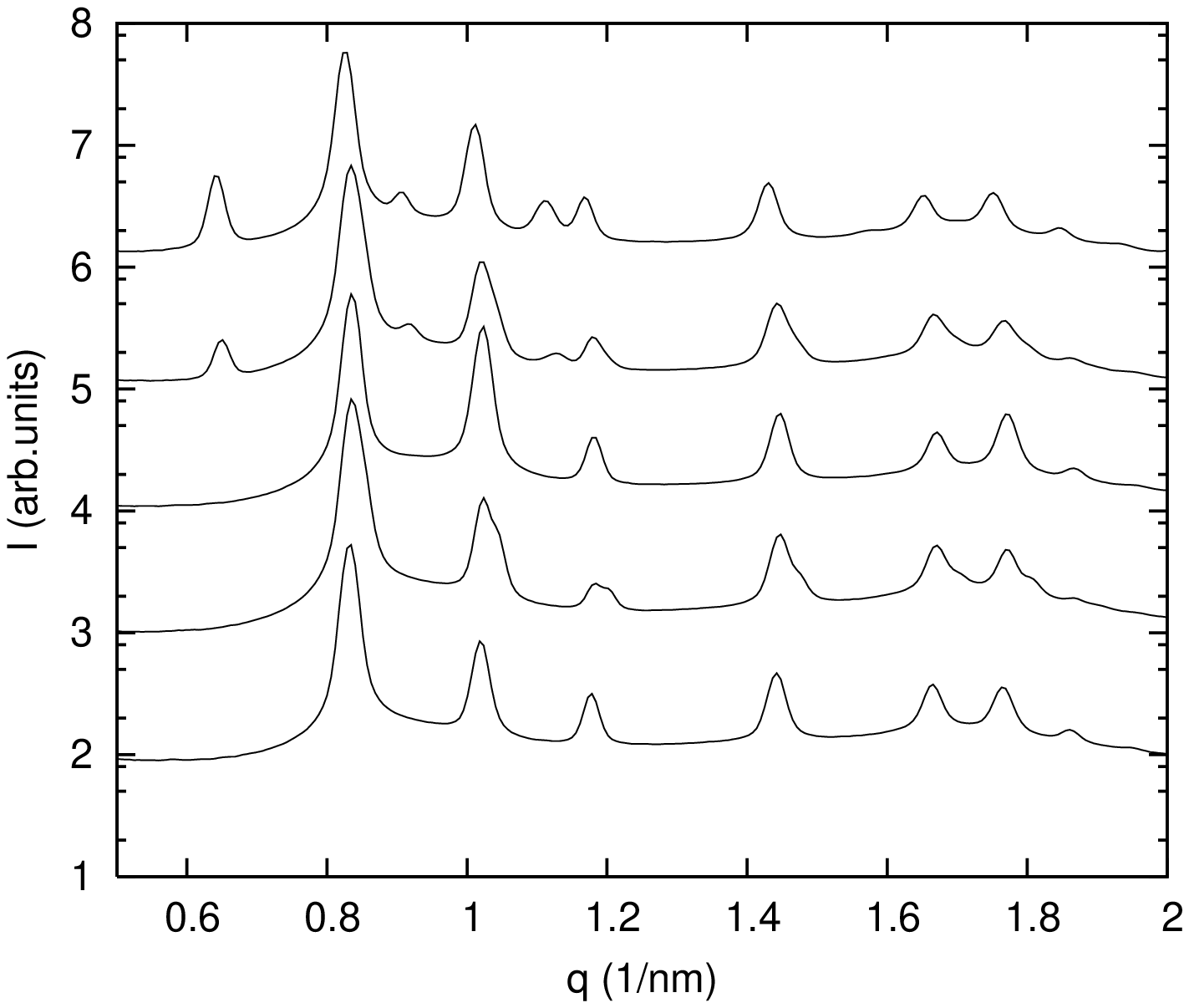}
(b)\includegraphics[width=5.5cm,angle=-90]{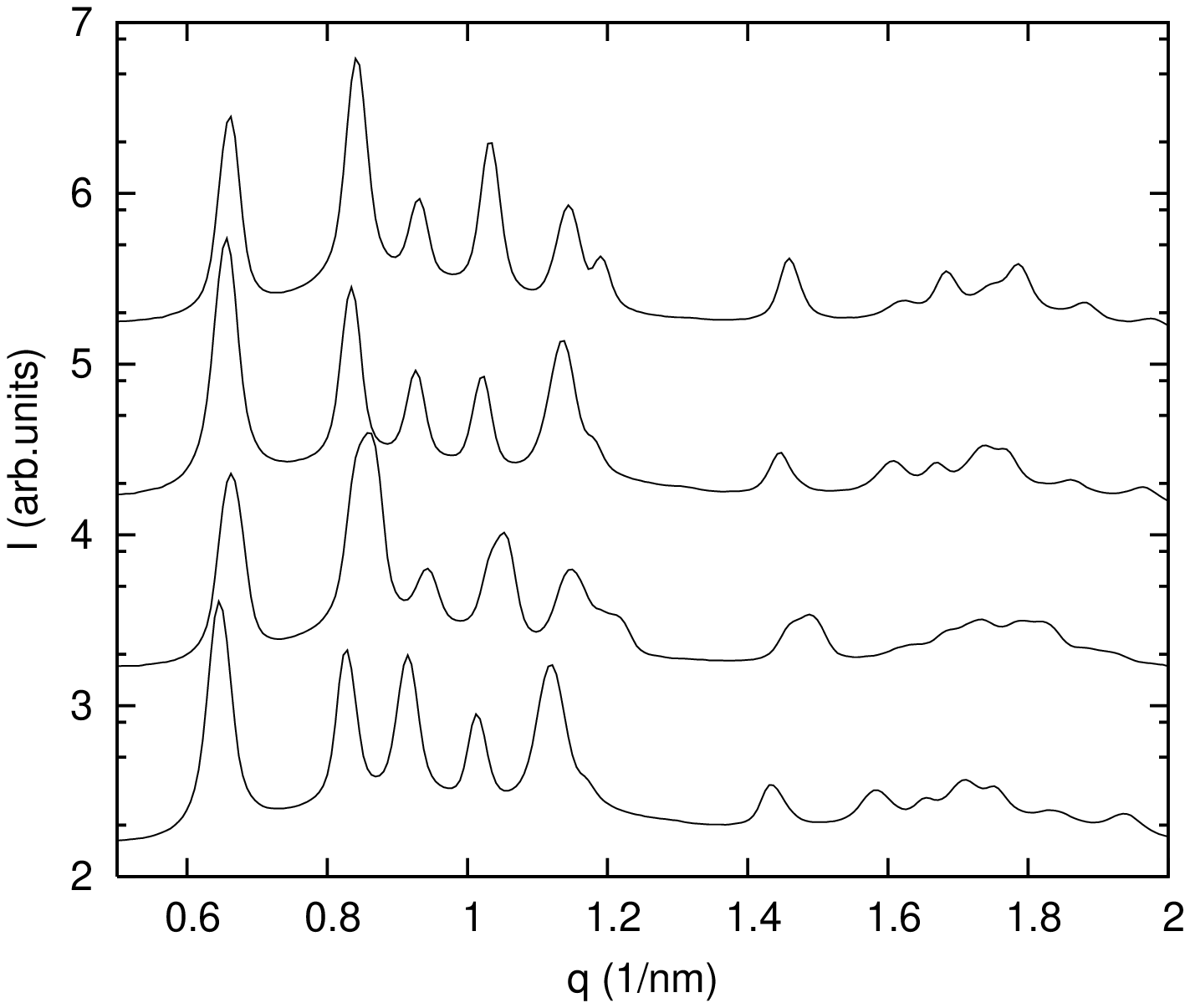}
\caption{SAXS patterns of the cubic phase with lysozyme. NaCl was not added. Concentration of lysozyme was increased from 0\% (bottom) to 4.0\% (top) by 1.0\% in (a), and from 5.0\%(bottom) to 8.0\%(top) by 1.0\% in (b).}\label{lyso0-8}
\end{figure}

We added lysozyme to buffer solutions to see how incorporated lysozyme changed the cubic phase structures. NaCl was not added yet at this stage. Figure \ref{lyso0-8} shows the effect of lysozyme on the SAXS patterns. The patterns were changed dramatically depending on the lysozyme concentration $c_{\rm lyso}$. At concentrations up to $c_{\rm lyso}=2.0$\%, the patterns were the same as the one at $c_{\rm lyso}=0$\% and they showed $\rm Pn\bar{3}m$ structure. The lattice parameters remained the same as well. At $c_{\rm lyso}=3.0$\%, however, new peaks appeared at about $q=0.7\ {\rm nm^{-1}}$, $0.9\ {\rm nm^{-1}}$, and $1.1\ {\rm nm^{-1}}$ (Fig. \ref{lyso0-8}a), and the height of these peaks grew with $c_{\rm lyso}$ up to $c_{\rm lyso}=5.0$\% (Fig. \ref{lyso0-8}b). At $c_{\rm lyso}$ larger than 6.0\%, these peaks were kept more or less constant in height (Fig. \ref{lyso0-8}b).

These three new peaks had spacing with the ratio of $\sqrt{2}$, $\sqrt{4}$, $\sqrt{6}$, which are indexed as (110), (200), (211) reflections from a cubic phase of space group $\rm Im\bar{3}m$. The appearance of $\rm Im\bar{3}m$ structure induced by lysozyme agrees with literature \cite{Larsson83b,Larsson96}. The lattice parameters of the new peaks were estimated as $13.7\pm 0.1$ nm. These lattice parameters did not change much with the lysozyme concentration.

\subsection{Effect of lysozyme with NaCl}

\begin{figure}
(a)\includegraphics[width=5.5cm,angle=-90]{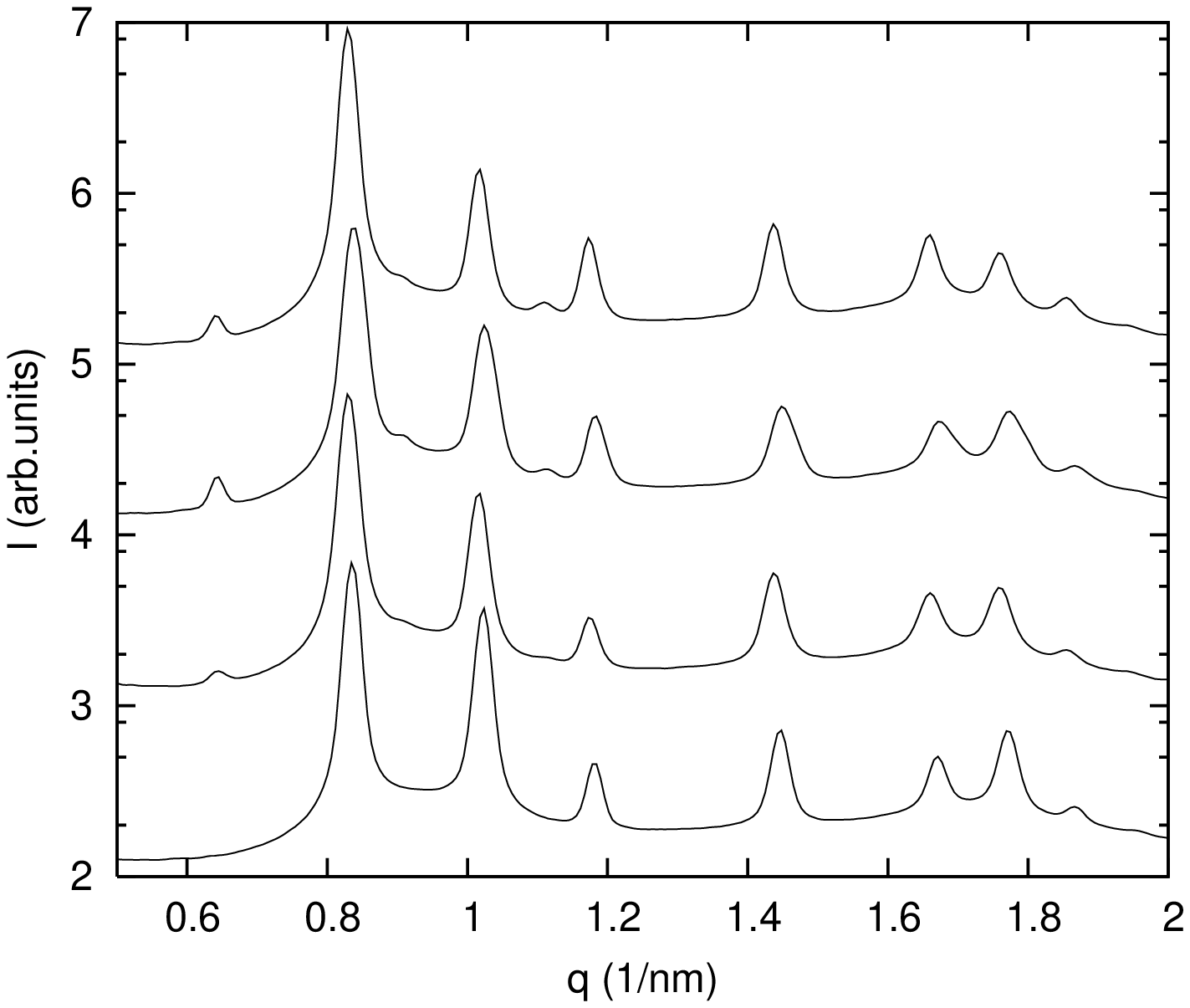}
(b)\includegraphics[width=5.5cm,angle=-90]{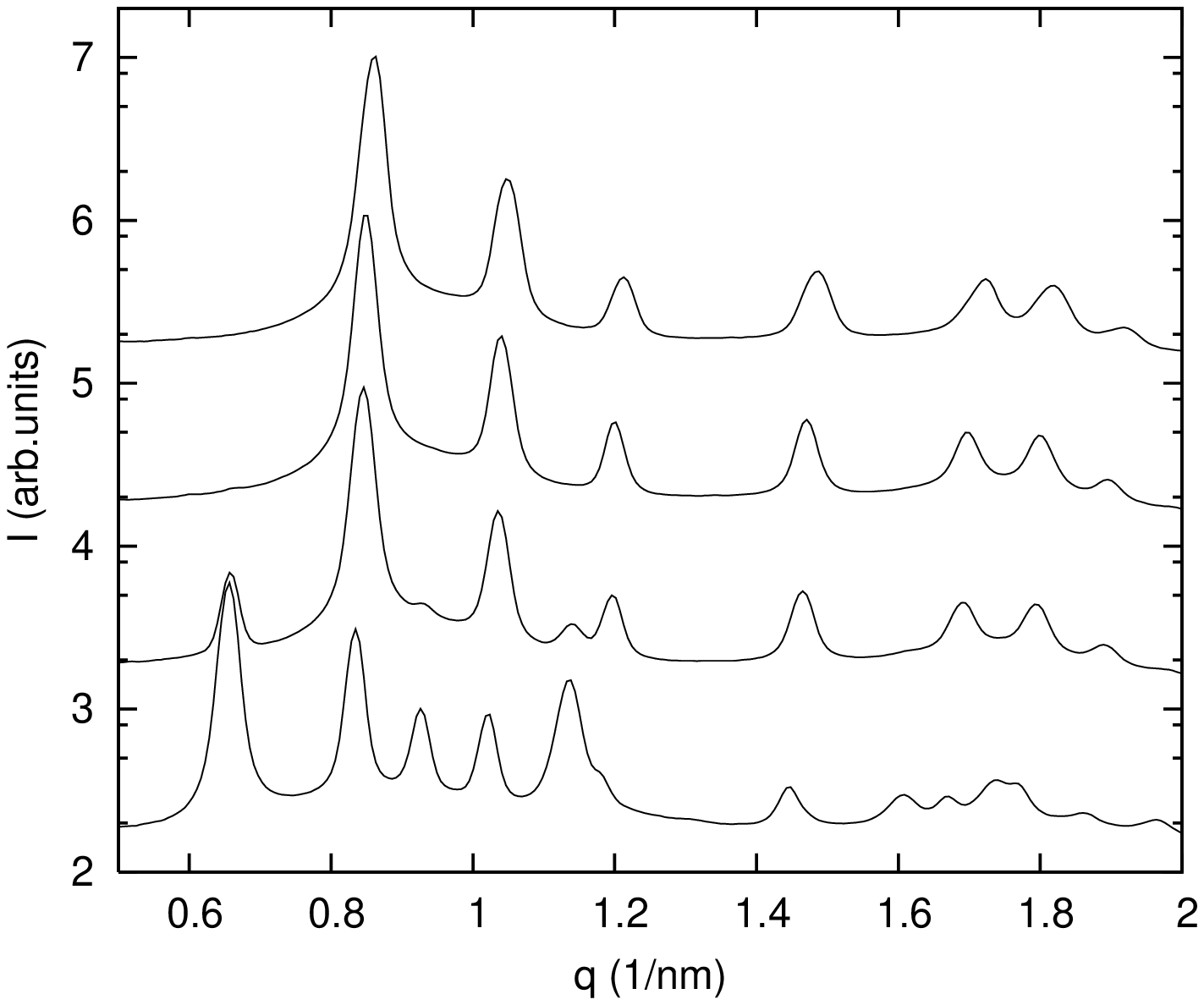}
\caption{SAXS patterns of the cubic phase with lysozyme plus NaCl. Lysozyme concentration was fixed at 2.0\% in (a) and at 7.0\% in (b). NaCl concentration is increased from 0 M (bottom) to 0.3 M (top) by 0.1 M.}
\label{lyso_NaCl}
\end{figure}

Lysozyme molecules are positively charged at pH 4.5. Ions like Cl$^{-}$ screen the surface charges of lysozyme, and therefore reduce repulsive interaction between lysozyme molecules. Figure \ref{lyso_NaCl}a shows the effect of NaCl at a fixed lysozyme concentration of $c_{\rm lyso}=2.0$\%. Without NaCl, the SAXS pattern at $c_{\rm lyso}=2.0$\% showed the $\rm Pn\bar{3}m$ structure (bottom one in fig.~\ref{lyso0-8}a). With increasing NaCl concentration $c_{\rm NaCl}$, a new peak appeared at about $q=0.7\ {\rm nm^{-1}}$ which is the same position as peaks appearing in Fig.~\ref{lyso0-8} at high lysozyme concentration ($c_{\rm lyso}\geq 3.0$\%). This suggests that the new structure belonged to $\rm Im\bar{3}m$, and that screening of the repulsive interaction between lysozyme molecules brought about it at lower lysozyme concentration than in the solutions without NaCl. In the range of $2\%\lesssim c_{\rm lyso}\lesssim 4$\%, the addition of 0.1-0.2 M NaCl either induced or strengthened peaks from the $\rm Im\bar{3}m$ structure.

Next we added NaCl to solutions containing $7.0$\% lysozyme (fig.~\ref{lyso_NaCl}b). At $c_{\rm lyso}=7.0$\%, the pattern already showed peaks from the $\rm Im\bar{3}m$ in addition to those from the $\rm Pn\bar{3}m$ without NaCl (the pattern at the bottom in Fig.~\ref{lyso_NaCl}b). With increasing NaCl concentration, however, the peaks from the $\rm Im\bar{3}m$ decreased as shown in Fig.~\ref{lyso_NaCl}b. At the NaCl concentration more than 0.2 M, the $\rm Im\bar{3}m$ structure disappeared completely. This tendency was observed in the region of high concentrations of lysozyme ($c_{\rm lyso}\gtrsim 4$\%) and NaCl ($c_{\rm NaCl}\gtrsim 0.2$M).

\begin{figure}
\includegraphics[width=6cm,angle=-90]{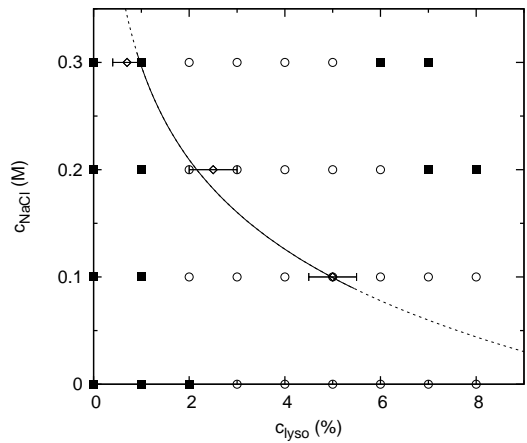}
\caption{A diagram of the structure observed by SAXS together with the solubility of lysozyme in the cubic phase. Solid squares represent the $\rm Pn\bar{3}m$ structure and open circles the coexistence of $\rm Pn\bar{3}m$ and $\rm Im\bar{3}m$. The open diamonds represent the position of crystal solubility of lysozyme in the cubic phase. The solid and dotted lines are guide to eye.}
\label{diagram}
\end{figure}

Figure \ref{diagram} shows a diagram of the cubic phase structure depending on lysozyme and NaCl concentrations. On the left side of the diagram, the $\rm Pn\bar{3}m$ structure was dominant. The $\rm Im\bar{3}m$ structure appeared around the middle of the diagram. Note that the undersaturated solutions (under the solid curve) could induce the $\rm Im\bar{3}m$ structures. On the upper right of the diagram, the $\rm Im\bar{3}m$ structure disappeared and only the $\rm Pn\bar{3}m$ structure remained.

The samples above the solubility curves were supersaturated. Though we did not observe lysozyme crystals during our SAXS experiments, the crystallization was proceeding. Accordingly the structure of the cubic phase could be affected by the ongoing crystallization process and change with time. In fact, crystal growth was observed in the samples after SAXS experiments. In preliminary experiments, we also observed that the SAXS patterns changed with time, where the change tended to follow the order of $\rm Pn\bar{3}m+Im\bar{3}m$ $\to$ $\rm Pn\bar{3}m$. 

\section{Discussion}
\subsection{Degeneracy of the $\rm Pn\bar{3}m$ and $\rm Im\bar{3}m$ structures}
Many lipid-water systems are now known to form bicontinuous cubic phases \cite{Luzzati67,Turner92,Templer98,Larsson83b,Larsson79,Nylander00}. It is also known \cite{Hyde96} that three cubic structures with the space groups of $\rm Ia\bar{3}d$, $\rm Pn\bar{3}m$, and $\rm Im\bar{3}m$ are most common and are formed on the basis of triply periodic minimal surfaces (TPMS), called respectively ``G (gyroid)'', ``D (diamond)'' and ``P (primitive)'' surfaces. The midplane of lipid bilayers corresponds to these surfaces. 

These three minimal surfaces are related with each other by the Bonnet transformation, which does not change the intrinsic geometry of the surface. This means that the associated curvature energy of these minimal surfaces is degenerate \cite{Hyde96}, though in real situation the degeneracy is removed due to the deviation from the ideal minimal surfaces. The thickness of the bilayers is probably the most important since it causes packing frustration of lipid molecules \cite{Turner92,Templer98,Templer98b}. Even if the degeneracy is removed in real situations, the energy of minimal surfaces may be reversed with small perturbations. 
 
The Bonnet transformation predicts that the ratio of lattice parameters of these minimal surfaces is $a_G:a_D:a_P=1.576:1:1.279$ \cite{Hyde84,Hyde96}, where subscripts represent each surface. Our results showed that the ratio was $a_{\rm Im\bar{3}m}/a_{\rm Pn\bar{3}m}=1.28$, which agrees well with this theoretical prediction. This suggests for the first time to our knowledge that the transition from the $\rm Pn\bar{3}m$ structure to the $\rm Im\bar{3}m$ structure of the fully hydrated monoolein cubic phase induced by lysozyme is in fact the transition of the cubic phase based on the D minimal surface to the cubic phase based on the P minimal surface related by Bonnet transformation. 

\subsection{Free volume for lysozyme molecules}
Our results reconfirmed the previously reported phenomenon \cite{Larsson83b,Larsson96}, that is, lysozyme molecules induce the transition of the monoolein cubic phase, from the $\rm Pn\bar{3}m$ structure to the $\rm Im\bar{3}m$ one. We also confirmed already that these two structures are closely related by the Bonnet transformation of the underlying minimal surfaces. In this section, we will discuss the possible driving force for this transition in terms of the frustration of lysozyme confined in the narrow channels.

\subsubsection{Calculations of volumes}
Bicontinuous cubic phases of monoolein consist of water channels with 4-fold junctions ($\rm Pn\bar{3}m$) or 6-fold junctions ($\rm Im\bar{3}m$). Lysozyme molecules, which are very soluble in water, likely stay in the channels. This has been confirmed using Raman scattering and calorimetric measurements \cite{Larsson96}. 

If we envision a bilayer membrane as two monolayers draped on the underlying minimal surface in parallel with a thickness $l$, the (dimensionless) volume of the bilayer may be calculated as
\begin{equation}
v_m = \frac{1}{a^3}\int_{-l}^l\dd l \int\dd A^l
\label{eq:1}
\end{equation}
where $\dd A^l$ is the differential area element on the parallel surface, which is expressed as \cite{Schwarz01}
\begin{equation}
\dd A^l = \dd A (1+Kl^2).
\end{equation}
where $K$ is the Gaussian curvature of the surface, and $\dd A$ the differential area element on the minimal surface. Then eq. (\ref{eq:1}) is calculated as
\begin{eqnarray}
\nonumber
v_m &=& \frac{1}{a^3}\int_{-l}^l\dd l \int\dd A (1+Kl^2) \\
\label{eq:2}
&=& 2\sigma_0\frac{l}{a} + \frac{4\pi}{3}\chi\left(\frac{l}{a}\right)^3,
\end{eqnarray}
where the Gauss-Bonnet theorem,
\begin{equation}
\int K\dd A = 2\pi\chi
\end{equation}
has been used. $\sigma_0=A/V^{2/3}$ is the dimensionless surface-to-volume ratio of the minimal surface, where $A$ is the surface area and $V$ the volume of the unit cell. $\chi$ is the Euler characteristic per unit cell. Then (dimensionless) volume of the water channel in the unit cell is
\begin{equation}
v_w=1-v_m.
\end{equation}
The values of $\sigma_0$ and $\chi$ for D and P minimal surfaces are summarized in table \ref{parameters}. 

\begin{table}[th]
\begin{center}
\caption{The Euler characteristic $\chi$ and the dimensionless surface to volume ratio $\sigma_0$ for D and P triply periodic minimal surfaces.}
\label{parameters}
\begin{tabular*}{\columnwidth}{@{\extracolsep{\fill}}cccc}\hline
TPMS type	&Space group	&$\chi$	&$\sigma_0$	\\ \hline
D			&$\rm Pn\bar{3}m$			&-2		&1.9189		\\
P			&$\rm Im\bar{3}m$			&-4		&2.3451		\\
\hline
\end{tabular*}
\end{center}
\end{table}

\begin{table}
\caption{The volume of water channels $v_w$, the volume available for the center of lysozyme molecules $v_a$, and the confinement factor $\alpha$ as the ratio of $v_w$ to $v_a$, $\alpha\equiv v_w/v_a$.}
\label{parameters2}
\begin{center}
\begin{tabular*}{\columnwidth}{@{\extracolsep{\fill}}ccccc}\hline
Space group	&$a$ (nm)	&$v_w$			&$v_a$			&$\alpha$	\\\hline
$\rm Pn\bar{3}m$		&10.7		&$0.41\pm0.01$	&$0.04\pm0.01$	&$8-14$		\\
$\rm Im\bar{3}m$		&13.7		&$0.44\pm0.01$	&$0.09\pm0.01$	&$4-6$		\\
\hline
\end{tabular*}\\
Note that $v_m$, $v_w$, and $v_a$ are in the unit of $a^3$.
\end{center}
\end{table}

\subsubsection{Available volume for lysozyme}
It is clear that lysozyme molecules impose some frustration on the $\rm Pn\bar{3}m$ structure so that $\rm Pn\bar{3}m$$\to$$\rm Im\bar{3}m$ transition occurs. In order to estimate the degree of the frustration, we use a simple concept of available volume for lysozyme \cite{Tanaka04} as follows. 

In the narrow water channels, available space for the motion of lysozyme is restricted. The center of molecules cannot go toward membranes beyond their radius. This limitation determines the volume available for lysozyme molecules $v_a (<v_w)$. Therefore, in dilute limit, their translational entropy should be calculated using $v_a$ instead of $v_w$. The difference between $v_a$ and $v_w$ is negligible if and only if a container is a bulk, but becomes significant if the container is nanosized. 

The effective concentration $c_{\rm eff}$ in the water channels is then estimated by $c_{\rm eff}=\alpha c$, $\alpha\equiv v_w/v_a$ where $c$ is the bulk concentration and $\alpha$ is the confinement factor. We estimated $v_a$ by assuming the radius of lysozyme as $r=1.7$ nm and the thickness of the monolayer as $l=1.75$ nm \cite{Caffrey96}, so the distance between the minimal surface and the position from there available for the center of lysozyme molecules is $l+r=3.45$ nm. We did not include the effect of hydration water on both lysozyme molecules and lipid molecules in this crude approximation, though it certainly has some role. Thus $v_a$ may be calculated using eq. (\ref{eq:2}) with $l\to l+r$. The values of $v_w$, $v_a$ and $\alpha$ are listed in table \ref{parameters2}. The value of $\alpha_{\rm Im\bar{3}m}=4-6$ agrees well with our previous estimation \cite{Tanaka04}, where $\alpha_{\rm Im\bar{3}m}$ was roughly calculated as the ratio of the volumes of two bipyramids which represented the space of an available volume and a water channel.    

As shown in Table \ref{parameters2}, whereas the volumes of the water channel $v_w$ for both structures are almost the same, the available volume $v_a$ of the $\rm Im\bar{3}m$ structure is more than twice as large as that of the $\rm Pn\bar{3}m$ structure. This is mainly because the $\rm Im\bar{3}m$ structure has 6-fold junctions where the water channel becomes wider than the 4-fold junctions in the $\rm Pn\bar{3}m$ structure.  

At the appearance of $\rm Im\bar{3}m$ structure, the concentration of lysozyme was about 3\% which corresponds to the molecular number density in solutions of about $1\times10^{-3}\ \rm nm^{-3}$. This means that there were about 0.5 molecule per $\rm Pn\bar{3}m$ cell ($a^3v_w=500\ \rm nm^{3}$) and about 1 molecule per $\rm Im\bar{3}m$ cell ($a^3v_w=1100\ \rm nm^3$). In both cases, therefore, we assume that the interaction between lysozyme molecules can be ignored if NaCl is not added. So the solutions are treated as ideal in the calculations that follow.   

The chemical potential of lysozyme molecules in confined geometry may be written as \cite{Tanaka04}
\begin{equation}
\mu = \mu_i + k_BT\ln \alpha
\end{equation}
where $\mu_i$ is the chemical potential in ideal bulk solutions. Thus the gain in the chemical potential when the transition from the $\rm Pn\bar{3}m$ structure to $\rm Im\bar{3}m$ one occurs is estimated as
\begin{equation}
\Delta \mu = k_BT\ln(\alpha_{\rm Im\bar{3}m}/\alpha_{\rm Pn\bar{3}m})\simeq -(0.3\sim1.3)k_BT
\label{eq:3}
\end{equation}
if we use the values of $\alpha$ in Table \ref{parameters2}. Note that this is the value per lysozyme molecule. At $c_{\rm lyso}\simeq 3$\%, it is approximately the free energy gain per cell since a cell contains approximately one molecule at this concentration.

Several authors have estimated the difference in the free energy between $\rm Pn\bar{3}m$ and $\rm Im\bar{3}m$ phase \cite{Turner92,Templer98b,Schwarz00}. It seems that there is a consensus about the smallness of the difference, since the curvature energy between the three minimal surfaces (G, D, and P) is in principle degenerate. Schwarz and Gommper \cite{Schwarz00} calculated the free energy density of several types of cubic phase. From their calculation of the free energy density (Fig. 2 in \cite{Schwarz00}) the free energy difference between $\rm Pn\bar{3}m$ and $\rm Im\bar{3}m$ cubic phase is estimated as about 2$k_BT$ per unit cell when the bending rigidity $\kappa=10k_BT$ and the spontaneous curvature $c_0=1/(6l)$, though the values of parameters are different from those of monoolein, $\kappa\simeq 3k_BT$ and $c_0\simeq 1/l$ \cite{Templer00}.

So far there is no direct measurement or calculation for the free energy difference between $\rm Pn\bar{3}m$ and $\rm Im\bar{3}m$ phase of monoolein-water system. As above, however, it is reasonably expected that the free energy to form the new cubic phase is compensated by the entropic gain for the lysozyme molecules shown in eq. (\ref{eq:3}).

\subsection{Structural transitions induced by moderate attraction between lysozyme molecules}
The SAXS patterns shown in Fig. \ref{lyso_NaCl} indicate that the addition of NaCl to lysozyme solutions tends to induce the $\rm Im\bar{3}m$ structure when both $c_{\rm lyso}$ and $c_{\rm NaCl}$ are not high. Note that when lysozyme and NaCl concentrations were high (on the upper right region in fig. \ref{diagram}), the structure observed was $\rm Pn\bar{3}m$, not $\rm Im\bar{3}m$, and we will consider it later in the next section. 

Lysozyme molecules at pH 4.5 are positively charged since the isoelectric point of lysozyme is about 11. Added chloride ions screen the surface charges of lysozyme, then induce effective attraction between lysozyme molecules. The source of the attraction is complex, probably the combination among the van der Waals force, hydrophobic interactions, hydrogen bonding, ion-bridging, and so forth. Irrespective of the source of the attraction, any attractively interacting molecules have the tendency to aggregate temporarily and/or permanently, or in other words, to move in concert, which was observed by dynamic light scattering \cite{Rosenberger95,Tanaka99}. Therefore the details of the effective attraction does not affect our discussion below.

It is natural to assume that these attractively interacting molecules induce more frustration in narrow water channels than those without attraction, since the collective motion of molecules is disturbed by membranes. We suggest that this attraction-induced frustration leads to the structural transition that occurs at lower lysozyme concentration than the solutions without attraction. 

\begin{figure}

\raisebox{4.8cm}{(a) }\includegraphics[width=6.5cm,angle=0]{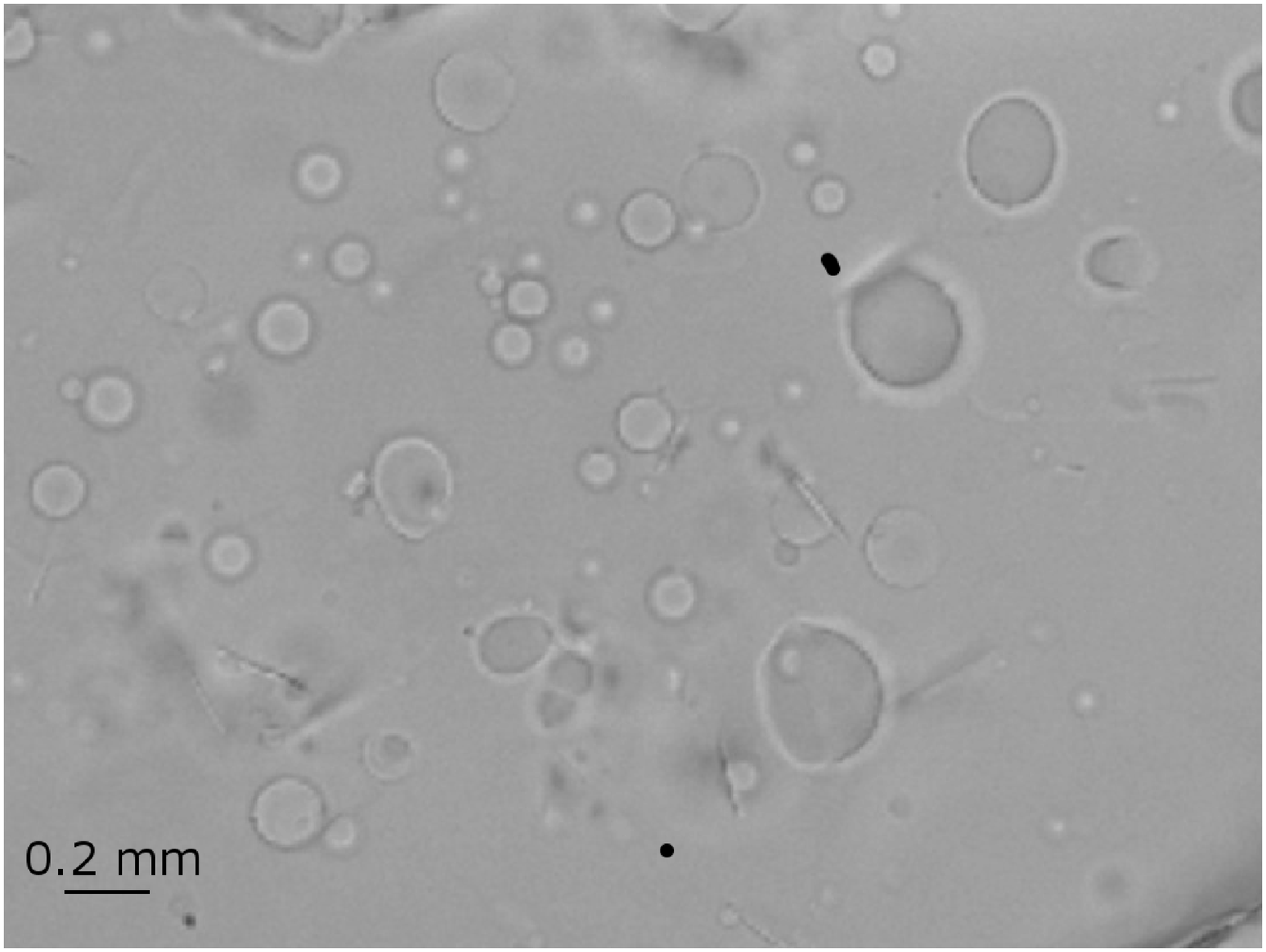}
\raisebox{4.8cm}{(b) }\includegraphics[width=6.5cm,angle=0]{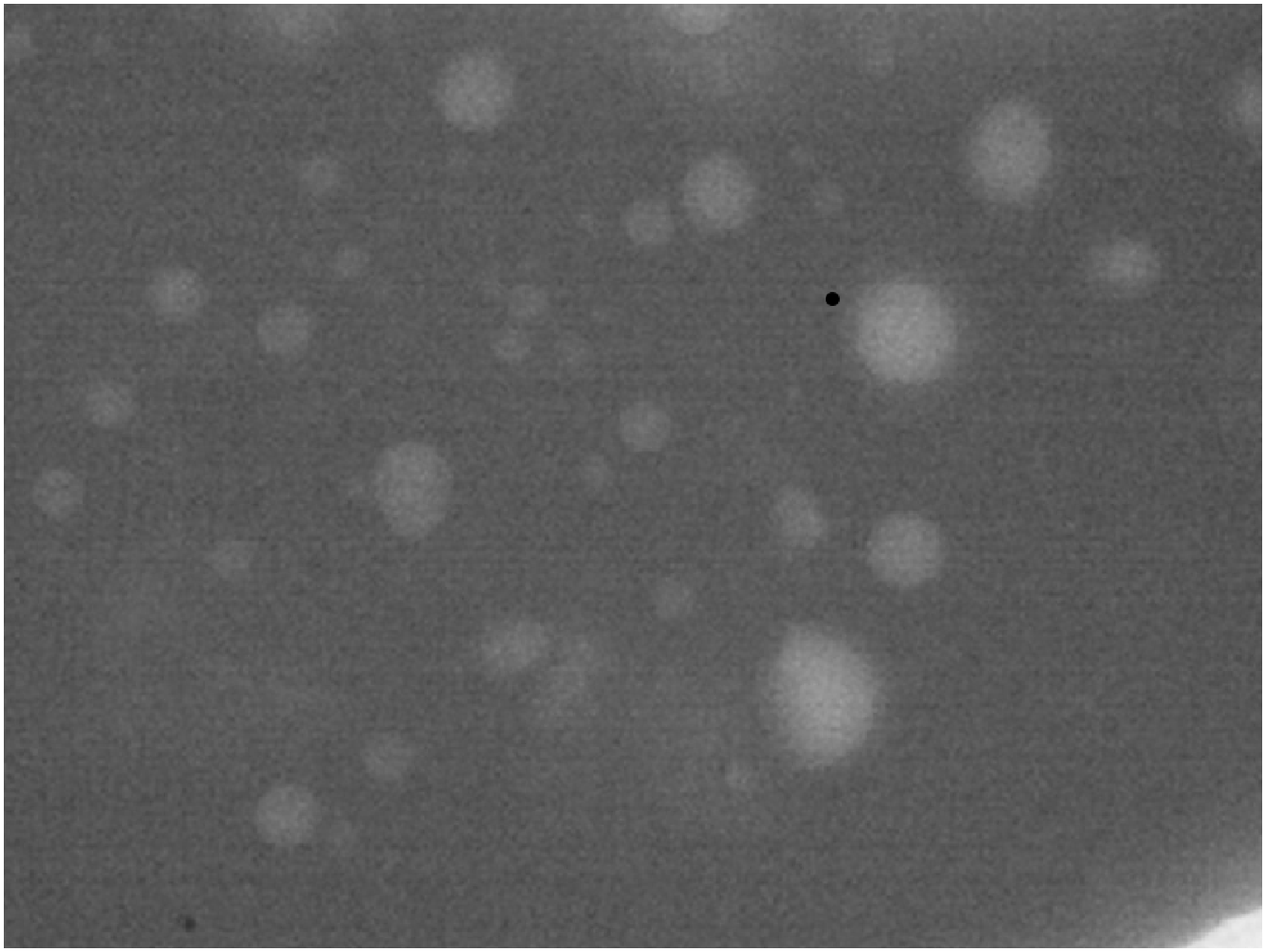}
\caption{Photographs of separated droplets taken with a bright field microscope (a) and with a fluorescence microscope (b). Lysozyme molecules were dyed with eosin Y. The fluorescence image shows that the droplets are filled with highly concentrated lysozyme solution. Around droplets is the transparent cubic phase. }
\label{droplets}
\end{figure}

\subsection{Crystallization, phase separation , and structural transitions}
Now we would like to discuss why $\rm Im\bar{3}m$ structure disappeared at the highest concentrations of lysozyme and NaCl. At concentrations above the solubility, crystallization proceeds. When nucleation takes place and protein crystals start to grow in the cubic phase, they probably destroy the surrounding, soft lipid structures until the protein concentration in the solution phase reaches equilibrium solubility of crystals. Then the frustration in the cubic phase is released on some level and the structure can turn back into the $\rm Pn\bar{3}m$. 

This scenario works well at $c_{\rm NaCl}=0.3$M, where the structure is $\rm Pn\bar{3}m$ on the solubility limit (Fig.~\ref{diagram}). It also works at $c_{\rm NaCl}=0.1$M, where the observed structure was the mixture of $\rm Pn\bar{3}m$ and $\rm Im\bar{3}m$ on the solubility, and no disappearance of the $\rm Im\bar{3}m$ phase observed. At $c_{\rm NaCl}=0.2$M, however, whereas the structure on the solubility was the mixture between $\rm Pn\bar{3}m$ and $\rm Im\bar{3}m$, the disappearance of the $\rm Im\bar{3}m$ was observed. Though it looks inconsistent with the scenario, we would like to point out that the peaks from $\rm Im\bar{3}m$ on the solubility ($c_{\rm lyso}\simeq2.0$\%) at $c_{\rm NaCl}=0.2$M shown in Fig. \ref{lyso_NaCl}(a) were much lower than those observed at higher concentrations, which suggests that the portion of $\rm Im\bar{3}m$ structure is small there. Moreover, at the next point to the solubility ($c_{\rm lyso}$=1\%), there was no $\rm Im\bar{3}m$ phase observed. Therefore, the results at $c_{\rm NaCl}=0.2$M shown in Fig. \ref{diagram} may agree with the above scenario within experimental errors. Further study will be needed to verify this.

Another mechanism to release the frustration from the cubic phase is the phase separation. Figure \ref{droplets} shows an example of the phase separation, sometimes observed in the samples, where lysozyme molecules are expelled from the cubic phase into droplets. The highly concentrated (almost glassy) state of lysozyme in the droplets was confirmed by fluorescence microscopy (Fig. \ref{droplets}b). These droplets thus serve as a reservoir, and lysozyme molecules are removed from the cubic phase into the droplets to release the frustration. The formation of the droplets is so irreproducible so far that we could not include them in the diagram shown in Fig. \ref{diagram}. However, the droplets tended to appear at high concentrations of lysozyme and NaCl. 
\vspace{5mm}

\section{Conclusions}
As proposed in \cite{Tanaka04}, the driving force of the enhanced crystallization in the cubic phase is the frustration in the water channels, where lysozyme molecules are highly confined. In this study we demonstrated that the same frustration drives the cubic phase into the structural transition. The frustration is induced mainly by the presence of lysozyme, but also by the attractive interaction between lysozyme molecules. 

The frustration removed by the transition from $\rm Pn\bar{3}m$ structure to the $\rm Im\bar{3}m$ one was estimated using a simple model of confinement, where translational entropy of lysozyme molecules was considered. The free energy gain due to the transition was estimated about $k_BT$ per lysozyme molecule. The smallness of the gain suggested the smallness of the free energy cost to transform the structure from $\rm Pn\bar{3}m$ to $\rm Im\bar{3}m$. We also showed experimentally that the two structures were related by the Bonnet transformation of the underlying minimal surfaces.   

The disappearance of $\rm Im\bar{3}m$ structure on the upper right corner of the diagram (Fig.\ref{diagram}) suggests that either crystallization or phase separation proceeds in the cubic phase. Expelling lysozyme into crystals or droplets formed by the phase separation removes the frustration, thus the cubic phase turns back into the $\rm Pn\bar{3}m$ structure which is stable when protein concentration is low. 

Water channels provided by lipidic cubic phases work as a nano-space to enhance crystallization of globular proteins. However, the present study revealed that the same frustration imposed by the cubic phase on the confined protein molecules can sometimes lead to the structural transition of the cubic phase, so as to expand the water channels. To use the lipidic cubic phase as a nanoscale container for effective protein crystallization, the interplay among lipid, protein, and crystallizing agent must further be understood.

\begin{acknowledgments}
The authors thank Dr. A. Toda for the critical reading of the manuscript and Dr. K. Inoue for his support in conducting SAXS experiments. The SAXS experiments were performed at the BL40B2 in the SPring-8 with the approval of the Japan Synchrotron Radiation Research Institute (Approval No. 2005B0095).
\end{acknowledgments}

\end{document}